**Ultrasonic investigation of the interaction hydrogen-dislocations in copper crystals.**

A. Moreno-Gobbi[1], G. Zamir[2] and J. A. Eiras[3]

[1]Instituto de Física - Facultad de Ciencias (UDELAR) - Iguá 4225 - CEP 11400 - Montevideo – Uruguay.
[2] Physics Department - Nuclear Research Center - Negev - Beer-Sheva 84190 (P.O.Box 9001) – Israel.
[3]Departamento de Física, UFSCar. Rod. Washington Luiz km 235, CEP 13565–905, Sao Carlos, Brazil.

Correspondent author: Ariel Moreno-Gobbi - moreno@fisica.edu.uy
Facultad de Ciencias - Universidad de la República
Iguá 4225 - 11400 - Montevideo - URUGUAY
Tel.: (+598-2) 525 8624
Fax: (+598-2) 525 0580

## Abstract
In this paper we present experimental data of ultrasonic velocity and attenuation obtained in a high purity crystalline sample of cooper hydrogenated by gaseous charge. The sample is oriented in the <111> crystallographic direction and aged for this work in three stages between 64 and 97 days. The results indicate that the hydrogen is mainly segregated at the dislocation core, inhibiting the Hydrogen Snoek-Köster relaxations verified at earlier aging stages. Despite of this, a contribution to viscosity in the kink-chain resonance is provided by the mobile hydrogen in the dislocations core by its side movement along the dislocation line. At temperatures at which the hydrogen begins to freeze in the lattice the geometrical kinks find a gradual increase on the hindering of their movements along dislocation lines, becoming immobile when the hydrogen is completely frozen in the crystal, anchoring the dislocations in short loops. Although the viscosity associated with the mobile hydrogen is removed, the resonance associated with geometrical kinks is not completely cancelled. The interaction hydrogen-dislocation can be fully described in terms of kinks in dislocations.

## 1. Introduction
Dislocations in crystalline metals can develop thermal and geometrical kinks. Thermal kinks are produced by pairs kink-antikink. Each pair is immediately annihilate unless an applied force keep them separate. In high purity crystals the effect of the phonons is to control the speed at which the kinks move along the dislocation driven by an external force [1]. The kink picture has proven to be the best approach to explain the internal friction and ultrasonic attenuation of cold-worked samples of high purity FCC metals [1]. At 10 MHz to 50 MHz frequencies the two attenuation peaks known as Niblet-Wilks and Bordoni peaks (hereafter denominated NWP and BP, respectively) are present in cold-worked high purity copper crystals, superposed to a dislocation related background, which were well described in terms of the kink picture [2]. While the NWR and BR relaxations are considered to be produced by the mechanism of kink pair formation with kink diffusion in the phonon field (KPF), the dislocation background is attributed to the resonance of the oscillating kink-chain with a kink density $\propto T$ [2].

In this picture each relaxation is associated to logarithmic decrement and modulus defect given in Equation 1 and 2 respectively,

$$\delta = \Delta_{Bi}.\left\{(\omega\tau)^2 / \left[1+(\omega\tau_i)^2\right]\right\} \tag{1}$$

$$\Delta M/M_o = \Delta_{Bi}.\left\{1/\left[1+(\omega\tau_i)^2\right]\right\} \tag{2}$$

Where $\Delta_{B(i)}$ and $\tau_i = B_{3(i)}.\left[(k_B T)^{3/2}.\exp\left(H_{k(i)}/k_B T\right)\right]$, $H_{k(i)}$ are respectively the relaxation intensity, the relaxation time and the activation enthalpy for each relaxation ($i$ = 1, 2), $k_B$ is the Boltzmann's constant and $\omega$ the angular frequency of the wave.
The relaxations are superimposed to the overdamped resonance of the kink chain, being the logarithmic decrement and the modulus defect described respectively by

$$\delta \propto N(T).\omega.L.B(T) \propto B_o.\ \omega.L^4.T^2 \tag{3}$$

and

$$\Delta M/M_0(T) \propto N(T).L^2 \propto L^2.T \tag{4}$$

Where $B(T)$ is the damping coefficient, $L$ the mean dislocation length between anchoring points and $N(T)$ the kink density of the kink chain conformed with geometrical kinks and those created below a critical temperature $T_c = \left(H_k/k_B\right).\ln\left[2a(1+N_0)^{1/2}/AL\right]$, produced at zero applied stress [2]. In this equation $a$ is the interatomic distance in the dislocation line direction, $N_0$ is the number of geometrical kinks in the dislocation line, $H_k$ = enthalpy of creation of one kink, $k_B$ is the Boltzmann's constant, and $A$ is a constant [2]. For high purity copper it holds that $B(T) \propto T$ [2] and, as we shown from experimental results, $N(T) \propto T$ [2].
The presence of hydrogen atoms in the crystal, although do not change the above picture, it can change the attenuation spectrum due to the interaction of hydrogen atoms with the dislocations. Since the relaxations are produced by the kink pair formation, the presence of hydrogen surrounding the dislocation must strongly affect these processes. When the sample is hydrogen-charged, the attractive interaction between dissolved hydrogen and dislocations induces high levels of hydrogen segregation close to the dislocation lines [3]. For the resonance background, produced by the kink-chain resonance, we can reasonably assume that the effect of the mobile hydrogen on the kink-chain is that of increasing the viscosity coefficient, although the temperature dependence remains unaltered. The main consequences of equations (1) to (4) remain applicable, but with different values for $H_k$, $B_0$ and $L$. As Seeger pointed out [1, 4] screw dislocations are responsible for BR while edge dislocations (60°) are responsible for NWR. Because of their hydrostatic stress field, the interaction with hydrogen is stronger for edge dislocations than for screw ones [5], indicating that different effects of hydrogen should be expected for each relaxation peak (NBR or BR).
In this paper we present measurements of phase velocity and attenuation of ultrasonic RF pulses, in a hydrogenated crystalline copper sample aged between 64 and 97 days, in order to study the interaction of diffused hydrogen with kinks in dislocations, and their comparison with those made at early stages of aging and published elsewhere [6].

## 2. Experimental procedure

High purity crystalline copper ((residual resistivity ratio (RRR) = 1500)) sample, in the form of a 10 mm side cube, was cut with two faces perpendicular to the <111> crystalline direction. The specimen, previously oriented by the Laue X-ray method, was 5.7 % deformed by compression in the direction <111>. After that it was charged with hydrogen during 72 hours at 0.2 MPa and 573 K. Further details of the experimental setup were described elsewhere [6].

For this work, ageing and ultrasonic measurements on the sample were accomplished in three stages. First, the sample was aged at room temperature and atmospheric pressure during 64 days. The entire process consists of three previous stages [6], which are: stage 1, immediately after the hydrogen-charging process was completed, the sample was aged for 1 h at room temperature and pressure, and under vacuum, the simple was cooled down from room temperature to 125K and then heated up to 350K at a rate of 1.5K/min; stage 2, after 26h ageing the second ultrasonic measurement was taken by heating the sample to 350K and then immediately cooling it down to 95K; stage 3, after 72h ageing, the last measurement was taken by cooling the sample down from 350K to 105K.

Measurements were realized cooling the sample from 294 K to 50 K and then warming until room temperature. Second, the sample was aged by 95 days prior to a heating up to 380 K, measurements were taking cooling the sample to 50 K and then cooling up to 335 K. Third, after the last measurements, the sample was aged for 2 days at room temperature after stage 2 and measurements were taken in a cooling-warming cycle.

Ultrasonic attenuation and phase velocity were simultaneously measured using the conventional pulse-echo method [7], with ultrasonic equipment MATEC and a JANIS closed cycle refrigerator, the whole experiment was automated in the laboratory. A quartz X-cut transducer of 10 MHz fundamental frequency was bonded with Nonac Stopcock grease to a <111> face of the sample, in order to generate longitudinal waves. The ultrasonic phase velocity was computed as $v = 2d/t$, where $d$ is the sample thickness and $t$ the elapsed time for a pulse round trip (transit time). The transit time was accurately determined with the Pulse-Echo-Overlap technique (PEO) [9].

The measurements were performed in the temperature range 50 K to 373 K, in cooling-heating cycles with the sample temperature varying at a rate of 1.5 K/min.

With the purpose to extract all effects other than those related to the dislocation contribution, the logarithmic decrement $\delta$ and modulus defect $\Delta M/M_0$ were obtained from experimental data with the following equations [7]:

$$\delta = 0.115(\alpha - \alpha_0) / f(\text{MHz}) \quad (5)$$

$$\Delta M / M_0 = (M_0 - M) / M_0 = (v_0^2 - v^2) / v_0^2 \quad (6)$$

Where $f$ (MHz) is the wave frequency expressed in megahertz; $\alpha_0$ the attenuation and $v_0$ the velocity of ultrasonic pulses measured in a dislocation-free sample (background); $\alpha$ the attenuation and $v$ the velocity of ultrasonic pulses measured in the sample. Both $\alpha_0$ and $v_0$ were measured in a <111> non-deformed copper sample, irradiated with γ-rays until total pinning of dislocations [8]. We assume that the temperature dependence of the dislocations free attenuation and velocity on a deformed sample is the same as that in undeformed samples [2]. Also corrections due to pulse path length variation with temperature were introduced [7].

**3. Experimental results and discussion**

In Figure 1 we present the ultrasonic attenuation (circles) and velocity (triangles) measured in a cooling-heating temperature sequence starting at 300 K, in the cold worked crystalline copper sample, hydrogen charged and aged at room temperature during 64 days. For comparison, the inset shows the ultrasonic attenuation for a high purity copper measured in an hydrogen free [2] and 3 days aged [6] hydrogenated samples, where the respective dislocation background were subtracted from the experimental data.

**Figure 1**

The general behaviour of the attenuation *vs.* temperature curves in Figure 1 can be divided in two different regions of temperature 300 K to 122 K and 122 K to 50 K. In the cooling run the attenuation shows an approximately linear decrease of 11.5 % between 300 K and 122 K, and below 122 K it shows an abrupt decrease of 33 % until 50 K is reached (below 50 K it seems to be independent of the temperature). In the heating run, started at 50 K, the attenuation remains nearly constant until 122 K and then presents a quick increase with the temperature until 175 K, and then it continues increasing in a lower linear rate at higher temperatures. Simultaneously, in the cooling run the associated velocity presents a change in slope around 122 K, while in the heating run between 122 K and 175 K it shows a marked decrease as the temperature increases. Figure 1 presents, as its more remarkable feature, the absence of noticeable attenuation peaks (NWR, BR or HSKR) in both cooling and heating branches.

For comparison, in the inset of Figure 1 we present attenuation *vs.* temperature obtained by us in a high purity crystalline copper sample uncharged [2] and in a hydrogen charged sample aged by three days [6]. The resonance background was subtracted from original data in both curves, in order to show more clearly the relaxation components. In the uncharged sample the NWR and BR can be clearly identified, while in the hydrogen charged and aged sample the presence of two HSKR is noticeable [6].

The absence of NWR and BR observed in Figure 1 could indicate that the hydrogen is still near the dislocation lines, inhibiting these peaks; while the absence of the two HSKR also indicates that the hydrogen atoms could not be in the dislocations surroundings.

In order to explain the behaviour shown in figure 1, we assume that the hydrogen atoms are located inside the dislocation core. Aiming to discuss only the dislocation-hydrogen contribution to the ultrasonic attenuation and velocity, in Figure 2 we present the logarithmic decrement and modulus defect only for the cooling branch of figure 1, obtained using equations (1) and (2).

**Figure 2**

It is clearly observed in Figure 2, as the temperature deceases from room temperature, a continuous softening of the effective modulus between room temperature and 122 K, followed by an abrupt hardening. This can be understood considering that as the temperature decreases from room temperature, the hydrogen atoms gradually lose it mobility and therefore the kinks gradually experiment less viscosity. In this way the crystal undergoes a softening process as the temperature decreases, as is observed in Figure 2 between room temperature and 122 K. Around 122 K it begins a freezing process of hydrogen atoms in the crystalline lattice, which produces an effective anchoring of the dislocations in shorter loops than those existing at temperatures above 122 K. At temperatures near 180 K a small variation of the modulus (as a relaxation) of about 0.1 %, is observed, corresponding to the small peak observed in the logarithmic decrement at the same temperature, presented in the inset of Figure 2. This small peak is produced at temperatures near to the corresponding to the relaxation associated to edge dislocations observed in the inset of Figure 1, which can indicate that a small amount of these dislocations still remain with atoms of hydrogen in their neighbourhood.

The logarithmic decrement, which can only be attributed to a hydrogen-dislocation mechanism, is originated by the oscillation of the kink chain. In fact, there are only two possible viscous mechanisms: the relaxation by KPF and the resonance of the geometrical kinks chain. As it is well known the crystalline phonon field is the main source for the effective viscosity experimented by kinks in high purity crystals. The effect of hydrogen atoms located in the dislocations core (which cannot move easily out of the dislocation core) is to pinning them into smaller segments. In this sense the segments can present a length smaller than that required for the KPF mechanism, and the KPF is inhibited. On the other hand, in those dislocation segments pinned at different Peierls valleys, where geometrical kinks can move along the Peierls directions, an additional contribution is now provided by the dragging of the hydrogen atoms moving along the Peierls directions.

If the temperature is high enough, the geometrical kinks can drag the hydrogen atoms. In other words, the existence of dislocation resonance without relaxations indicates that they have only mobility along dislocation lines (that is in the Peierls directions) allowing the oscillation of the kink chain. When the hydrogen becomes less mobile, as the temperature decreases, the geometrical kinks present progressively lower mobility. Finally, as the hydrogen freezes in the crystalline lattice the geometrical kinks become immobile or oscillate only along shorter dislocation segments between hydrogen atoms. This last is the responsible for the existence of a finite attenuation at low temperatures.

In the heating run of Figure 1, when the temperature starts to increase from 50 K, the hydrogen atoms in the dislocation core inhibit the kink movement until the temperature of 122 K is reached, at which the hydrogen begins to acquire increasing mobility and the logarithmic decrement start to increase. At a temperature around 175 K, the dissociation of hydrogen from kinks occurs and finally the geometrical kinks can move freely again, restricted only by the presence of mobile hydrogen atoms, increasing the effective viscosity. As it is well known, in high purity copper samples (RRR ≅ 1400) not cold worked, hydrogen is practically immobile at 123 K [10], which is in concordance with our interpretation of the experimental behaviour in our cold worked samples, in terms of hydrogen-kink interaction. It is also well known that at 140 K hydrogen atoms become mobile and can be trapped by certain impurities [10], and above 170 K it dissociates from atoms of impurities [11]. From what was already commented, in our context it is apparent that the kinks in dislocations act as impurities in the crystalline lattice.

In Figure 3 we present the ultrasonic velocity and attenuation obtained after 95 days of aging and room temperature after the measurements corresponding to Figure 1. Previously to measurements the sample was heated to a temperature of 380 K.

**Figure 3.**

The results shown in Figure 3 can be explained assuming that the heating up to 380 K promoted the diffusion of the Hydrogen atoms out of the dislocations core to it vicinity, what would be the reason for the the resurgence of the two HSK peaks. In the cooling branch of Figure 3, from a temperature of 160 K it is observed the appearance of an abrupt increase of the attenuation accompanied by a simultaneous decrease in the associated wave velocity. This behaviour can be explained considering that, at 160 K, the dislocations undergo a depinning process from Hydrogen atoms. Hydrogen atoms surrounding the dislocations experiment the effect of three different mechanisms: KPF in dislocations, freezing in the crystal lattice and the attraction by edge dislocations (which are responsible for the NWR at 105 K in the hydrogen free sample). The last mechanism induce the repining of the dislocations by Hydrogen atoms, promoting a decrease of the dislocation loop lengths, what implies the reduction of attenuation and hardening of the elastic modules at lower temperatures.
In Figure 4 we present experimental data obtained on the sample aged at room temperature for 2 days after the measurements of Figure 3.

**Figure 4.**

It is worthwhile to note in Figure 4 that, after the process of Figure 3, the attenuation and velocity *vs.* temperature curves become similar to that observed in Figure 1. This indicates that hydrogen atoms returned to their positions deeper in the dislocations core, inhibiting again the dislocation-hydrogen relaxation peaks.
All our experimental data and analysis can be well explained in terms of hydrogen diffusion to the core of the dislocations, and the interaction of hydrogen atoms with kinks in dislocations.

## 4. Conclusions

Experimental data of ultrasonic velocity and attenuation obtained on an hydrogenated high purity crystalline copper sample oriented in the <111> crystallographic direction, aged in 64, 95 and 97 days were analysed. The results indicated that the Hydrogen Snoek- Köster relaxations verified at the early ageing stages are inhibited by the hydrogen segregation at the dislocations core after the charged sample is aged in 64 days at room temperature. KPF and HSK are inhibited, but the geometrical kink chain resonance is still observed, because geometrical kinks remain mobile in dislocation lines, and hydrogen atoms can move through the core of dislocation lines, providing a contribution to a certain kink-viscosity. At temperatures when hydrogen atoms freeze in the crystalline lattice, the geometrical kinks become immobile. This effect can explain the behaviour observed in both attenuation and elastic modulus at temperatures between 50 K and 140 K. This work brings strong experimental evidence of the validity of the kink picture for dislocation vibration and shows that the geometrical kinks can play the role of impurities in relation to the interaction with the hydrogen diffused in the crystal.

## Acknowledgements.

We thank PEDECIBA and ANII in Uruguay by financial support.

**FIGURE CAPTIONS**

**Figure 1.** Longitudinal ultrasonic attenuation (circles) and velocity (triangles) at 10 MHz measured in a cooling (closed)-heating (open) run starting at 300 K, in a cold worked crystalline copper sample charged with hydrogen and aged in 64 days at room temperature. Inset: ultrasonic attenuation for a high purity copper measured in a hydrogen free [2] and a 3 days aged [6] hydrogenated samples, where the respective backgrounds were subtracted from the experimental data.

**Figure 2**. Logarithmic decrement and modulus defect for the cooling branch of Figure 1, obtained using equations (1) and (2). Inset: details of the relaxation around 160 K.

**Figure 3.** Ultrasonic velocity and attenuation obtained in the sample aged during 95 days. Before measurements, the sample was heated up to 380 K and then cooled slowly until 350 K when the measurements started. Circles: cooling run, triangles: run.

**Figure 4.** Ultrasonic velocity and attenuation obtained in the sample 2 days after the measurements obtained for Figure 3, aged at room temperature. Circles: cooling run, triangles: heating run.

**FIGURES**

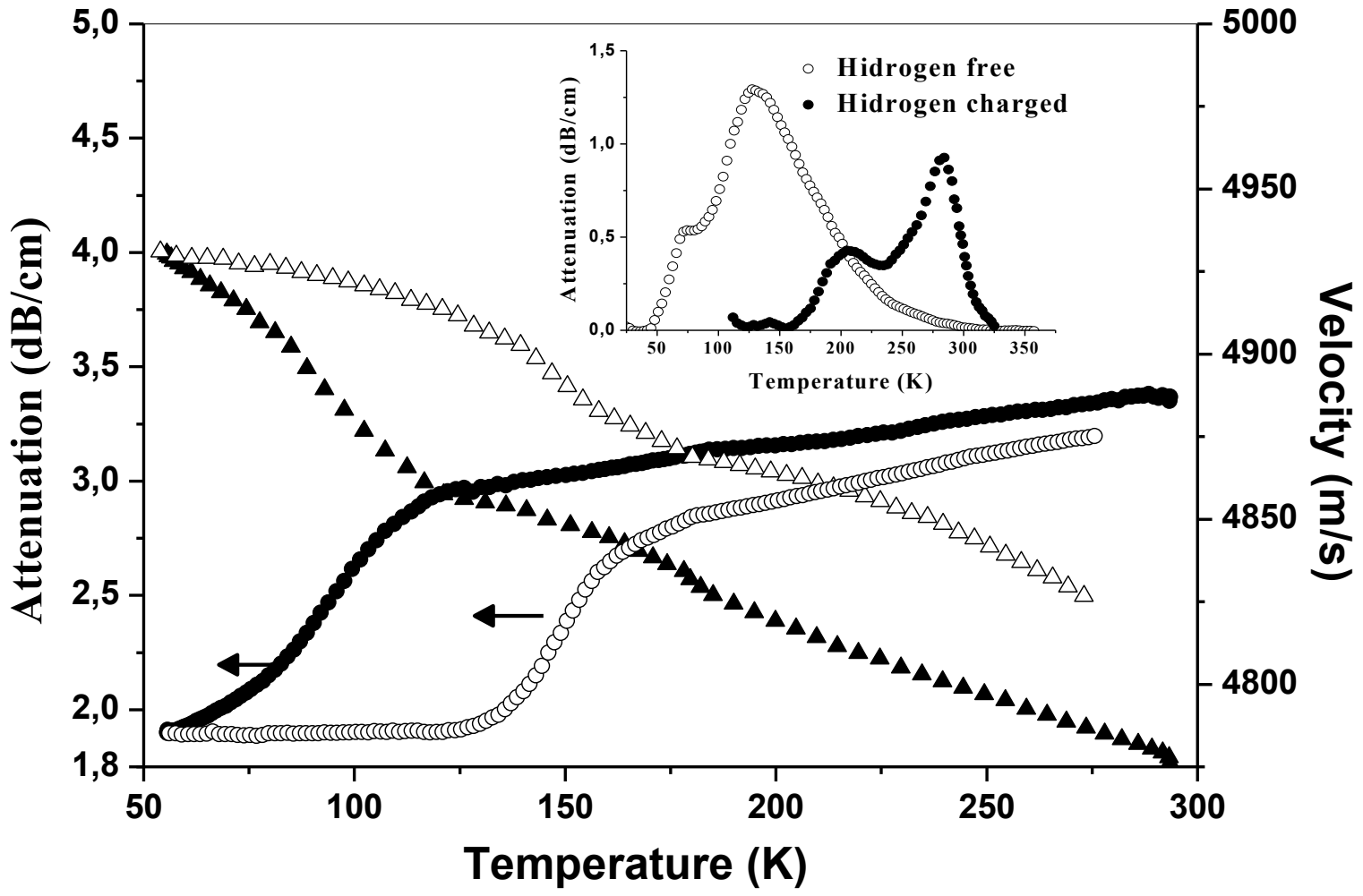

**Figure 1**

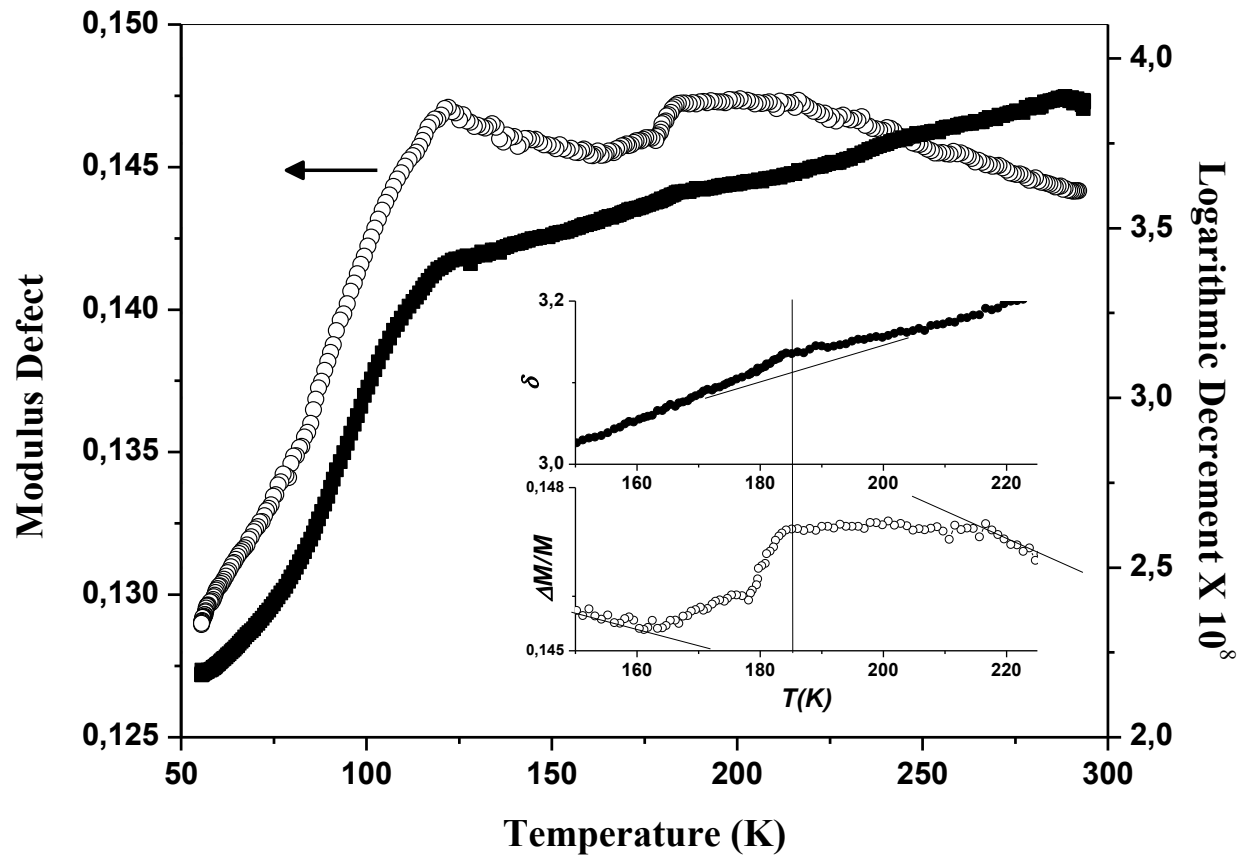

**Figure 2**

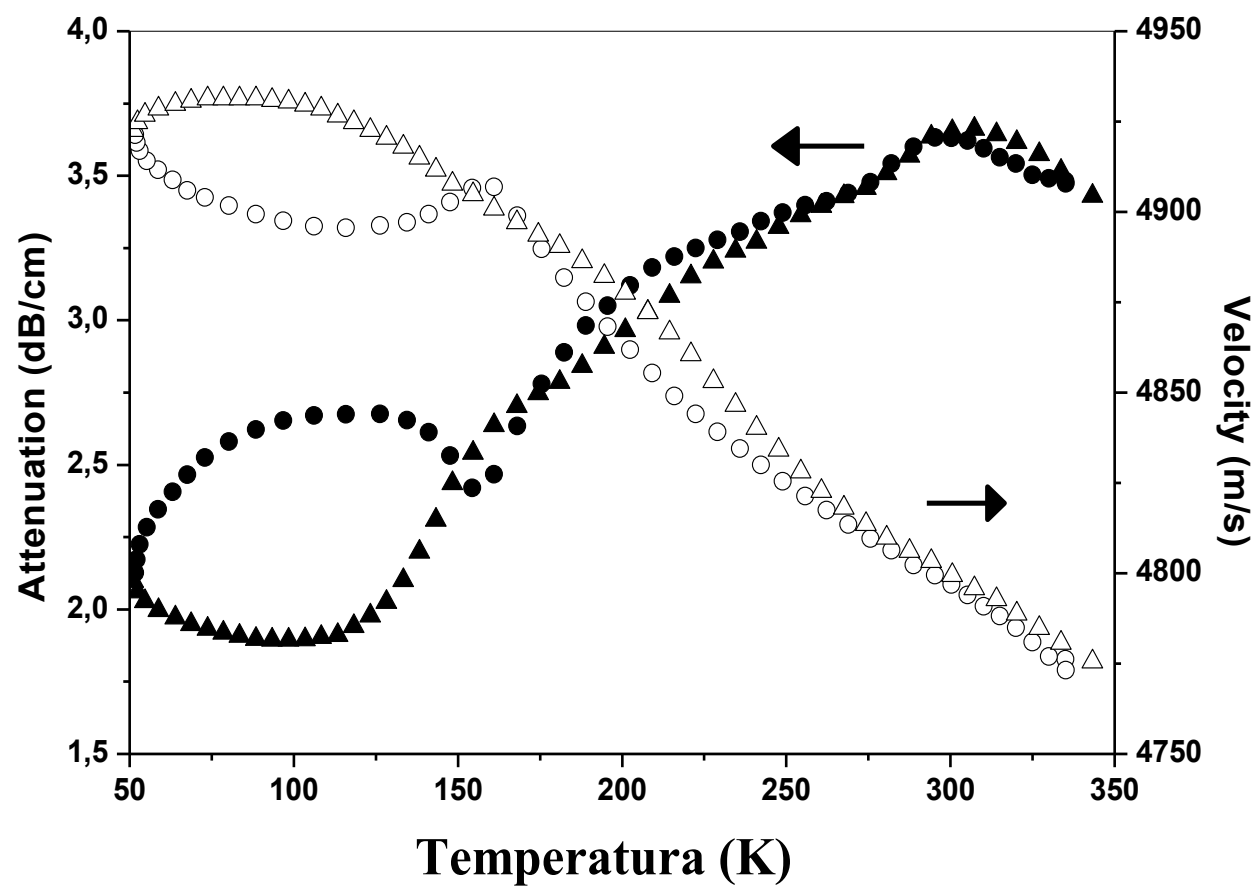

**Figure 3**

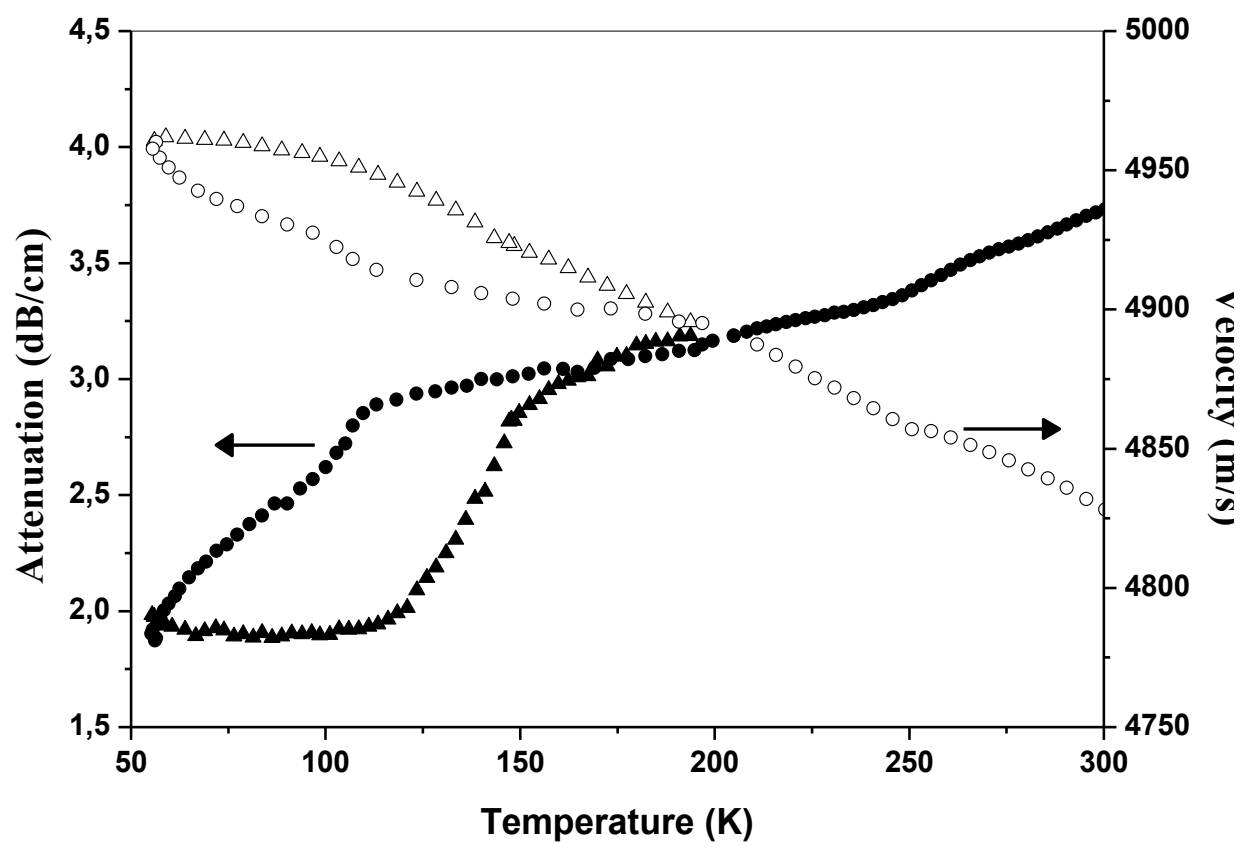

**Figure 4**